\documentclass[11pt]{article}
\pdfoutput=1
\usepackage{times}
\usepackage[centertags]{amsmath}
\usepackage{amssymb}
\usepackage{float}
\usepackage{algorithmic}
\usepackage{algorithm}

\newtheorem{theorem}{Theorem}
\newtheorem{proposition}{Proposition}[section]
\newtheorem{definition}{Definition}[section]

\newtheorem{lemma}{Lemma}[section]

\newtheorem{problem}{Problem}

\providecommand{\probab}[2]{\mathbb{P}_{#2}\left\{#1\right\}}

\providecommand{\binom}[2]{{#1\choose#2}}

\newenvironment{proof}[0]{\textit{Proof.} }{\hfill  $\blacksquare$ } 

\newenvironment{problem-env}[1]{
\vskip 3mm
\begin{problem} 
\textbf{#1}
}
{
\end{problem}
}

\newenvironment{neat-list}[3]{
  \vskip 1mm
  \begin{list}{}
  {\itemsep 2mm  \labelsep #1 \labelwidth #2 \leftmargin #3
   \partopsep 0mm \topsep 0mm \parsep 0mm  \itemindent 0mm}
}
{
  \end{list}
  \vskip 3mm
}

\newcommand{\weicnf}[3]{{\mathcal F}_{#3}^{#1,#2}}

\newcommand{\bnot}[1]{\overline{#1}}
\newcommand{\bor}{\vee}

\begin{document}
 
\title{A Fixed-Parameter Algorithm for Random Instances of Weighted $d$-CNF Satisfiability}

\author{
Yong Gao\thanks{Work supported by NSERC Discovery Grant RGPIN 327587-06} \\
Department of Computer Science \\
    Irving K. Barber School of Arts and Sciences \\
    University of British Columbia Okanagan \\
    Kelowna, Canada V1V 1V7 \\
    }
\maketitle

\begin{abstract}
We study random instances of the weighted $d$-CNF satisfiability problem (WEIGHTED $d$-SAT),
a generic W[1]-complete problem. %
A random instance of the problem consists of a fixed parameter $k$ and a random
$d$-CNF formula $\weicnf{n}{p}{k, d}$ generated as follows: 
for each subset of $d$
variables and with probability $p$,  a clause over the $d$ variables is selected
uniformly at random from among the $2^d - 1$  clauses that contain at least one 
negated literals. 

We show that random  instances of WEIGHTED $d$-SAT  can be solved
in $O(k^2n + n^{O(1)})$-time with high probability,
indicating that typical instances of
WEIGHTED $d$-SAT under this instance distribution are fixed-parameter tractable.   
The result also hold for random instances from the model $\weicnf{n}{p}{k,d}(d')$
where clauses containing less than $d' (1 < d' < d)$ negated literals are forbidden, and
for random instances of  the renormalized (miniaturized)  version of WEIGHTED $d$-SAT in
certain range of the random model's parameter $p(n)$.  This, together with our previous
results on the threshold behavior and the resolution complexity of unsatisfiable instances
of $\weicnf{n}{p}{k, d}$, provides an almost complete characterization of the typical-case
behavior of random instances of WEIGHTED $d$-SAT.      
  
\end{abstract}

\section{Introduction}
The theory of parameterized complexity and fixed-parameter algorithms is becoming an active research area in recent years \cite{downey99book,rolf06book}.
Parameterized complexity provides a new perspective on hard algorithmic problems, while
fixed-parameter algorithms have
found applications in a variety of areas such as artificial intelligence,
computational biology, cognitive modeling,
graph theory, and various optimization problems.

The study of the typical-case behavior of random instances of  NP-complete problems and coNP-complete problems such as satisfiability (SAT) and graph coloring has had much impact on our understanding of the nature of hard problems
as well as the strength and weakness of algorithms and well-founded heuristics
\cite{achlioptas01sharp,beame02efficiency,cheeseman91,cook97}.
Designing polynomial-time algorithms that solve random instances of NP-complete
problems under various random distributions has also been an active research area.  

In this work, we extend this line of research to intractable parameterized
problems. We study random instances of the weighted $d$-CNF satisfiability problem 
(WEIGHTED $d$-SAT), a generic W[1]-complete parameterized problem.
An instance of WEIGHTED $d$-SAT consists of a $d$-CNF formula $\mathcal{F}$ and a
fixed parameter $k > 0$. The question is to decide if there is a satisfying assignment
with Hamming distance $k$ to the all-zero assignment. A variant of 
WEIGHTED $d$-SAT is MINI-WEIGHTED $d$-SAT that asks if there is a satisfying assignment 
with Hamming distance $k\log n$ to the all-zero assignment.
 
We show that there is an  $O(k^2n + n^{O(1)})$-time  algorithm
that solves random instances of WEIGHTED $d$-SAT
with high probability for any $p(n) = \frac{c\log n}{n^{d - 1}}$. 
The result also hold for random instances from the more general model $\weicnf{n}{p}{k,d}(d')$
where clauses containing less than $d' (1 < d' < d)$ negated literals are forbidden, and for
random instances of MINI-WEIGHTED $d$-SAT with
the random model's parameter $p(n)$ being in a certain range. 
This, together with our previous
results on the threshold behavior and resolution complexity of unsatisfiable instances
of $\weicnf{n}{p}{k, d}$ in \cite{yong08aaai}, provides a nearly complete characterization of the typical-case
behavior of random instances of WEIGHTED $d$-SAT. To the best knowledge of the author, this is the first work in the literature on the fixed-parameter tractability 
of random instances of a W[1]-complete problem.   

The main result of this paper is  that instances from the
random distribution $\weicnf{n}{p}{k, d}$ (and its generalization $\weicnf{n}{p}{k,d}(d')$) 
of WEIGHTED $d$-SAT are ``typically" fixed-parameter tractable for any
$p = \frac{c\log n}{n^{d - 1}}$ with $c > 0$.   
\begin{theorem}
\label{th-wei2cnf-sat}
There is an $O(k^2n + n^{O(1)})$-time algorithm that with high probability, either 
finds a satisfying assignment of weight $k$ or reports that no such assignment exists for 
a random instance $(\weicnf{n}{p}{k,d}, k)$ of
WEIGHTED $d$-SAT for any $p = \frac{c\log n}{n^{d - 1}}$ with $c > 0$.
\end{theorem}

In the appendices, we show that the same algorithm can be extended to solve 
random instances from the  more general model $\weicnf{n}{p}{k, d}(d')$ and 
random instances of MINI-WEIGHTED d-SAT for certain range of the probability parameter
$p(n)$.
The next section contains necessary preliminaries and
a detailed description of the random model. In Section 3, we present
the algorithm W-SAT together with a discussion on its time complexity.
In Section 4, we prove that W-SAT succeeds with high probability for
random instances of WEIGHTED $d$-SAT.
In the last section, we discuss directions for future work.

\section{Preliminaries and Random Models for WEIGHTED $d$-SAT}

An instance of a \textit{parameterized decision problem} is a pair $(I, k)$ where
$I$ is a problem instance and $k$ is the problem parameter \cite{downey99book,rolf06book}.
Usually, the parameter $k$ either specifies the ``size" of the solution or
is related to some structural property of the underlying problem, such as the treewidth
of a graph.  
A parameterized problem is fixed-parameter tractable (FPT) if any instance $(I, k)$ of the problem can be solved in $f(k)|I|^{O(1)}$ time, where $f(k)$ is a computable function that
depends only on $k$. 
Parameterized problems are inter-related by parameterized reductions, resulting in a classification of parameterized problems into a hierarchy of complexity classes
$
FPT %
\subseteq W[1] \subseteq W[2] \cdots \subseteq XP.
$
It is believed that
the inclusions are strict and  the notion of completeness can be naturally defined
via parameterized reductions.

\subsection{Weighted CNF Satisfiability and its Random Model}
As with the theory of NP-completeness, the satisfiability problem plays
an important role in the theory of parameterized complexity.
A CNF formula (over a set of Boolean variables) is a conjunction of disjunctions of
literals. A $d$-clause is a disjunction of $d$-literals. A $d$-CNF formula is a CNF
formula that consists of $d$-clauses only.
An assignment to a set of $n$ Boolean variables is a vector in $\{\textrm{TRUE, FALSE}\}^n$.
The \textbf{weight} of an assignment is the number of variables that are set to
TRUE by the assignment. It is convenient to identify TRUE with $1$ and FALSE with $0$. Thus, an assignment can also be regarded as a vector in $\{0, 1\}^n$ and the weight of an assignment
is just its Hamming distance to the all-zero assignment.  

A representative $W[1]$-complete problem
is the following weighted d-CNF satisfiability (WEIGHTED d-SAT) problem:
\begin{problem-env}{WEIGHTED d-SAT}
\label{pr-w-d-sat}
\begin{neat-list}{3mm}{1.5cm}{2cm}
\item[\textbf{Instance:}\hfill] A CNF formula consisting of $d$-clauses,
         and a positive integer $k$.
\vskip 0.2cm
\item[\textbf{Question:}\hfill] Is there a satisfying assignment of weight $k$?
\end{neat-list}
\end{problem-env}

In \cite{marx05csp}, Marx studied the parameterized complexity of the more general
parameterized Boolean constraint satisfaction problem. One of the results of 
Marx (\cite{marx05csp}, Lemma 4.1), when applied to CNF formulas, is that any instance 
of WEIGHTED $d$-SAT can be reduced to at most $d^k$ instances each of which 
is a conjunction of clauses that contain at least one negated literal.  
Marx further proved that WEIGHTED $d$-SAT is W[1]-complete even when restricted to CNF
formulas that consist of clauses of the form $\bnot{x} \bor y$.

We use $G(n, p)$ to denote the Erd\"{o}s-Renyi random graph where $n$ is the number of vertices and $p$ is the edge probability \cite{bollobas01}. In $G(n, p)$, each of the possible $\binom{n}{2}$ edges appears
independently with probability $p$. A random hyper-graph $\mathcal{G}(n, p, d)$
is a hypergraph where each of the $\binom{n}{d}$ possible hyperedges appears independently with
probability $p$.
Throughout the paper, by ``with high probability" we mean that the probability
of the event under consideration is $1 - o(1)$.

We will be working with the following random model of WEIGHTED $d$-SAT, which is basically
similar in spirit to random CNF formulae with a planted solution studied in  
traditional (constraint)  satisfiability 
(See, e.g., \cite{alon97siam,feige01semi,flaxman-soda-03,gao07jair,krivelevich-soda-06,molloy02} 
and the references therein).
\begin{definition}
Let $X=\{x_1, \cdots, x_n\}$ be a set of Boolean variables and $p = p(n)$ be a function
of $n$. Let $k$ and $d$ be two positive constants.

We define a random model $\weicnf{n}{p}{k, d}$ for WEIGHTED d-SAT parameterized by $k$ as follows: To generate an instance $\mathcal{F}$ from $\weicnf{n}{p}{k, d}$, we first construct
a random hypergraph $\mathcal{G}(n, p, d)$ using $X$ as the vertex set. For each hyperedge
$\{x_{i_1}, \cdots, x_{i_d}\}$, we include in $\mathcal{F}$ a $d$-clause selected uniformly
at random from the set of $2^d - 1$ non-monotone $d$-clauses defined over the variables
$\{x_{i_1}, \cdots, x_{i_d}\}$. 
(A monotone clause is a clause that contains positive literals only).

\end{definition}

The model $\weicnf{n}{p}{k, d}$ can be generalized to $\weicnf{n}{p}{k, d}(d')$ as follows:
instead of from the set of non-monotone clauses, we select uniformly 
at random from the set of 
clauses over $\{x_{i_1}, \cdots, x_{i_d}\}$ that contain at least $d'$ negated literals.  
Note that $\weicnf{n}{p}{k, d}$ is just $\weicnf{n}{p}{k, d}(1)$. In the rest of this paper, we
will be focusing on $\weicnf{n}{p}{k, d}$, but will discuss how the algorithm 
and the results  can be adapted to $\weicnf{n}{p}{k, d}(d')$ in Appendix A.
 
Note that since monotone clauses are excluded,
the all-zero assignment always satisfies a random instance of $\weicnf{n}{p}{k, d}$ in the traditional sense.  On the other hand, in view of Marx's results we mentioned earlier in this
subsection, forbidding monotone clauses is not really a restriction.  As a matter of fact, our study
begins with a random model that doesn't  pose any restriction on the type of clauses that can appear in  a formula. Such a model, however,  turns out to be trivially unsatisfiable since unless the model parameter $p(n)$ is extremely small,  a random instance will contain more than $2k$ independent monotone clauses.

\subsection{Residual Graphs of CNF Formulas and Induced Formulas}
Associated with a CNF formula is its \textbf{residual graph} over the set
of variables involved in the formula. There is an edge between two variables
if they both occur in some common clause. The residual graph of a random instance
of $\weicnf{n}{p}{k, 2}$ is the random graph $G(n, p)$. The residual graph
of a random instance of $\weicnf{n}{p}{k, d}$ is the \textbf{primal graph}
of the random hypergraph $\mathcal{G}(n, p, d)$.

Let $\mathcal{F}$ be a $d$-CNF formula and $V \subset X$ be a subset of variables.
The \textbf{induced formula} $\mathcal{F}_V$ of $\mathcal{F}$ over $V$ is defined
to be the CNF formula $\mathcal{F}_V$ that consists of the following two types of clauses:
\begin{enumerate}
\item the clauses in $\mathcal{F}$ that only involve the variables in $V$;
\item the clauses of size at least $2$ obtained by removing any literal whose corresponding
 variables are in $X \setminus V$.
\end{enumerate}

\section{A Fixed-Parameter Algorithm for Instances of $\weicnf{n}{p}{k, d}$ }
In this section, we describe the details of the fixed-parameter algorithm designed for
random instances of $\weicnf{n}{p}{k, d}$ and show that its time complexity is
$O(k^2n + n^{O(1)})$. The results in this section and in the next section together establish
Theorem \ref{th-wei2cnf-sat}.

\subsection{General Idea}
We describe the general idea in terms of WEIGHTED 2-SAT. A detailed description
of the algorithm for $\weicnf{n}{p}{k, d}$ is given in the next subsection. The generalization
of the algorithm to the more general random model $\weicnf{n}{p}{k, d}(d')$ is presented
in Appendix A. %

The algorithm W-SAT considers all the variables
$x$ that appears in more than $k + 1$ clauses of the form $\bnot{x} \bor y$.
Any such variable cannot be assigned to TRUE.  
By assigning these forced variables to FALSE, we get a reduced formula. W-SAT then 
checks to see if the reduced formula can be decomposed into connected components of size at most $\log n$. If  no such decomposition is possible, W-SAT gives up. Otherwise,    
let $\{\mathcal{F}_i, 1\leq i\leq m\}$
be the collection of connected components in the reduced formula.
For each connected component $\mathcal{F}_i$,  use brute-force to find the set of
integers $L_i$ such that for each $k' \in L_i$, there is an assignment
of weight $k'$ to the variables in $\mathcal{F}_i$  that satisfies $\mathcal{F}_i$.

Finally, a dynamic programming algorithm is applied to 
find in time $O(k^2n)$ a collection of at most $k$ positive integers $\{k_{i_j}, 1\leq j\leq k\}$ such that
\begin{equation}
\label{eq-decomp}
\left\{\begin{array}{l}
k_{i_j} \in L_{i_j}, \textrm{ and } \nonumber \\
k_{i_1} + k_{i_2} + \cdots + k_{i_k} = k
\end{array} \right.
\end{equation}
Combining the weight-$k_{i_j}$ solutions to the subproblems indexed by $i_j$, a weight-$k$
solution can be found. If on the other hand, no such $\{k_{i_j}, 1\leq j\leq k\}$ can be found, we can safely report that the original instance has no weight-$k$ satisfying assignment.  

\subsection{Details of the Algorithm W-SAT} 
We first introduce the following concept that is essential to the algorithm:
\begin{definition}
Let $(\mathcal{F}, k)$ be an instance of WEIGHTED $d$-SAT where $\mathcal{F}$ is a 
$d$-CNF formula
and $k$ is the parameter.  Consider a variable $x$ and
a collection of subsets of variables $\mathcal{Y} = \{Y_i, 1\leq i\leq k\}$ where
$
  Y_i = \{y_{ij}, 1\leq j \leq (d - 1) \}
$
is a subset of $X \setminus \{x\}$.
We say that the collection $\mathcal{Y}$   \textbf{freezes} $x$ if
the following two conditions are satisfied:
\begin{enumerate}
\item  $Y_i \cap Y_j = \emptyset, \forall i, j$.
\item for each $1\leq i\leq k$, the clause
$
 \bnot{x} \bor y_{i1} \bor \cdots \bor y_{i(d-1)}
$
is in the formula $\mathcal{F}$.
\end{enumerate}
A variable $x$ is said to be \textbf{k-frozen} with respect to a subset of
variables $V$ if it is frozen by a collection of subsets of variables
$\{Y_i, 1\leq i\leq k\}$ such that $Y_i \subset V, \forall 1\leq i\leq k$.
A variable that is $k$-frozen with respect to the set of all variables is 
simply called a \textbf{$k$-frozen variable}.
\end{definition}
It is obvious that
a $k$-frozen variable cannot be assigned to TRUE without forcing  more than
$k$ other variables to be TRUE.
We also need the following concept to describe the algorithm:
\begin{definition}
Let $\mathcal{F}$ be a CNF formula. We use $L_{\mathcal{F}}$ to denote
the set of integers between $0$ and $k$ such that for each $k' \in L_{\mathcal{F}}$, there
is a satisfying assignment of weight $k'$ for $\mathcal{F}$.
\end{definition}

The algorithm W-SAT is described in Algorithm 1. We explain in the following
the purpose of the subroutine REDUCE().
The subroutine REDUCE($\mathcal{F}, U$)
simplifies the formula $\mathcal{F}$ after the variables in $U$
have been set to 0. It works in the same way as the unit-propagation based
inference in the well-known DPLL procedure for traditional satisfiability
search: It removes any clause that is satisfied by the assignment to the variables
in $U$; deletes
all the occurrences of a literal that has become FALSE due to the assignment;
and assigns a proper value to the variables
that are forced due to the literal-deletion. The procedure terminates
when there is no more forced variable.
It is easy to see the following lemma holds for the subroutine REDUCE():
\begin{lemma}
\label{lem-reduce}
REDUCE() never assigns TRUE to a variable.   
If $\mathcal{F}' = \textrm{REDUCE}(\mathcal{F}, U)$ is empty, then $\mathcal{F}$ has a
weight-$k$ satisfying assignment if and only if at least $k$ variables
have not been assigned by REDUCE().
\end{lemma}
 
\begin{algorithm}
\begin{algorithmic}[1]
\caption{W-SAT}
\label{algo-weighted-sat}
\REQUIRE An instance $(\mathcal{F}, k)$ of WEIGHTED $d$-SAT
\OUTPUT A satisfying assignment of weight $k$, or UNSAT, or FAILURE

\STATE  Find the set of $k$-frozen variables $U$ and assign
them to FALSE.
\STATE  Let $\mathcal{F}' = \textrm{REDUCE}(\mathcal{F}, U)$ be the \textbf{reduced formula}.
\STATE  Find the connected components $\{\mathcal{F}_1, \cdots, \mathcal{F}_m\}$
    of $\mathcal{F}'$.
\STATE If there is a connected component of size larger than $\log n$, return ``FAILURE".
\STATE Otherwise, for each connected component $\mathcal{F}_i$,
use brute force to find $L_{\mathcal{F}_i}$.
\STATE %
       Find a set of at most $k$ indices  $\{i_j, 1\leq j \leq k\}$
       and a set integers $\{k_{i_j}, 1\leq j \leq k\}$ such that
       $k_{i_j} \in L_{\mathcal{F}_{i_j}}$ and
 $
    \sum\limits_{j = 1}^{k} k_{i_j} = k.
 $
 Return ``UNSAT" if there is no such index set.
\STATE For each $\mathcal{F}_{i_j}$, use brute-force to find
  a weight-$k_{i_j}$ assignment to the variables in
  $\mathcal{F}_{i_j}$ that satisfies $\mathcal{F}_{i_j}$.
\STATE Combine the assignments found in the above to form a weight-$k$ satisfying
assignment to the formula $\mathcal{F}$.
\end{algorithmic}
\end{algorithm}

\subsection{Correctness and Time Complexity of W-SAT}
The correctness follows directly from the previous discussion. For the time complexity, we have the following
\begin{proposition}
\label{prop-time}
The running time of W-SAT is in $O(k^2n + n^{O(1)})$.
\end{proposition}
\begin{proof}
Since Lines 1 through 4, Line 5, and Line 7 together take $n^{O(1)}$ time, 
we only need to show that Line 6 can be done in $O(k^2n)$ time using dynamic programming.
Consider an integer $k$ and a collection $\{L_i, 1\leq i \leq m\}$ where
each $L_i$ is a subset of integers in $\{0, 1, \cdots, k\}$. We say that
an integer $a$ is \textbf{achievable} by $\{L_i, 1\leq i\leq m\}$
if there is a set of indices $I_a = \{i_j, 1\leq j\leq l\}$ such that
for each $i_j$, there is a $k_{i_j} \in L_{i_j}$  so that
$
 \sum\limits_{j = 1}^{l}k_{i_j} = k.
$
We call any such an index set $I_a$ a \textbf{representative set} of $a$.
The purpose of Line 6 is to check to see if the integer $k$ is \textbf{achievable},
and if YES, return a representative set of $k$. The Proposition follows from the follow lemma.
\end{proof}

\begin{lemma}
\label{lem-dynamic-prog}
Given a collection $\{L_i, 1\leq i \leq m\}$ and an integer $k$ where
each $L_i$ is a subset of integers in $\{0, 1, \cdots, k\}$, there is a dynamic
programming algorithm that finds a representative set of $k$ if $k$ is achievable, or
reports that $k$ is not achievable. It runs in time $O(k^2m)$.
\end{lemma}
\begin{proof}
Let $A(t) = \{(a, I_a): 0\leq a \leq k\}$ be the set of pairs
$(a, I_a)$ where $0\leq a\leq k$ is an integer achievable by 
$\{L_i, 1\leq i\leq t\}$ and $I_a$ is a representative set of $a$. 

Let $A(0) = \emptyset$. We see that $A(t + 1)$ consists of the pairs of the form
$((a + b), \overline{I}_a)$ satisfying
$$
\left\{ \begin{array}{l}
    (a, I_a)\in A(t), \\
    b \in L_{t + 1} \textrm{ such that }  b \leq k - a, \textrm{ and } \\
    \overline{I}_a = I_a \cup \{t\}.
   \end{array}
   \right.
$$
A typical application of dynamic programming builds $A(0), A(1), \cdots, \textrm{ and }A(m)$.
The value $k$ is achievable by $\{L_i, 1\leq i\leq m\}$ if and only if
there is a pair $(k, I_k)$ in $A(m)$. Since the size of $A(t)$ is at most $k$, the above algorithm runs in $O(k^2m)$ time.
\end{proof}

\section{Algorithm W-SAT Succeeds With High Probability}
\label{sec-analysis}
In this section, we prove that the
algorithm W-SAT succeeds with high probability on random instances
of $\weicnf{n}{p}{k, d}$.  Due to  Proposition \ref{prop-time}, we only
need to show that W-SAT reports ``FAILURE" with probability asymptotic
to zero. Recall that W-SAT  fails only when the reduced formula $\mathcal{F}'$ obtained in Line 2 has a connected component of size at least $\log n$.
The rest of this section is devoted to the proof of the following Proposition:
\begin{proposition}
\label{prop-conn-comp}
Let $\mathcal{F} = \weicnf{n}{p}{k, d}$ be the input random CNF formula
to W-SAT. With high probability,
the residual graph of the induced formula  $\mathcal{F}_V$ on $V$ decomposes into a collection of connected components of size at most $\log n$, where
$V$ is the set of variables that are not $k$-frozen.
\end{proposition}
\begin{proof}
Let $X = \{x_1, \cdots, x_n\}$ be the set of Boolean variables, and
let $U$ be the set of \textbf{$k$-frozen} variables so that  $V = X \setminus U$.
Since $p = \frac{c\log n}{n^{d - 1}}$ with $c > 0$, there will be many $k$-frozen
variables so that the size of $U$ is large.
If $U$ were a randomly-selected subset of variables, the proposition  is easy to prove. The difficulty in our case is that $U$
is not randomly-selected, and consequently
$\mathcal{F}_V$ cannot be assumed to be distributed in the same manner as the input
formula $\mathcal{F}$.

To get around this difficulty, we instead directly upper bound the probability $P^*$  
that the residual graph of $\weicnf{n}{p}{k, d}$ 
contains as its subgraph a  tree  $T$ over a given set $V_T$ of $\log n$ variables such that every variable $x \in V_T$ is not $k$-frozen. 
Since the variables in $\mathcal{F}_V$ are not $k$-frozen, an upper bound on
$P^{*}$ is also an upper bound on the probability that the residual graph of $\mathcal{F}_V$ contains as its subgraph a tree of the size $\log n$. We then use this upper bound  together
with Markov's inequality to show that the probability that the residual graph of $\mathcal{F}_V$ has a connected component of size at least $\log n$ tends to zero.

Let $T$ be a fixed tree over a subset $V_T$ of $\log n$ variables.  The difficulty
in estimating $P^*$ is that the event that the residual graph of $\weicnf{n}{p}{k, d}$ 
contains $T$ as its subgraph  and the event that no variable in $T$ is $k$-frozen are not independent of each other. To decouple the dependency, 
we consider the following two events:
\begin{enumerate}
\item $\mathcal{A}$: the event that the residual graph of  
   $\weicnf{n}{p}{k, d}$ contains the tree $T$ as its subgraph; and
\item $\mathcal{B}$: the event that none of the variables in $V_T$
  is $k$-frozen \textbf{with respect to $X \setminus V_T$}.
\end{enumerate}
Since by definition, being $k$-frozen with respect to a subset of variables
implies being $k$-frozen with respect to all variables, we have
\begin{equation}
\label{eq-2sat-conn-1}
P^* \leq \probab{\mathcal{A} \cap \mathcal{B}}{}.
\end{equation}
We now claim that
\begin{lemma}
\label{lem-2sat-independency}
The two events $\mathcal{A}$ and $\mathcal{B}$ are independent, i.e.,
\begin{equation}
\label{eq-2sat-conn-2}
\probab{\mathcal{A} | \mathcal{B}}{} = \probab{\mathcal{B}}{}
\end{equation}
\end{lemma}
\begin{proof}
Note that the event $\mathcal{A}$
depends only on those $d$-clauses that contain at least two variables in $V_T$ and that
the event $\mathcal{B}$ depends only on those $d$-clauses that
contain exactly one variable in $V_T$.
Due to the definition of the random model $\weicnf{n}{p}{k, d}$, the appearance
of a clause defined over a $d$-tuple of variables is independent from the appearance
of the other clauses. The lemma follows.
\end{proof}

Based on Equation (\ref{eq-2sat-conn-1}) and Lemma \ref{lem-2sat-independency}, we only
need to estimate $\probab{\mathcal{A}}{}$ and $\probab{\mathcal{B}}{}$.
The following lemma bounds the probability that a variable is not $k$-frozen.
\begin{lemma}
\label{lem-dsat-frozenp}
Let $x$ be a  variable and $W \subset X$ such that $x \in W$ and
$|W| > n - \log n $. We have
\begin{eqnarray*}
&&\probab{x \textrm{ is not } k\textrm{-frozen with respect to } W }{} 
  \leq O(1) \max(\frac{1}{n^{\delta}}, \frac{\log^2 n}{n})
\end{eqnarray*}
where $0 < \delta < \frac{c}{3(2^{d} - 1)(d - 1)!}$.
\end{lemma}
\begin{proof}
Let $N_x$ be the number of clauses of the form
$\bnot{x} \bor y_1 \bor \cdots \bor y_{d - 1}$ with
$\{y_1, \cdots, y_{d - 1}\} \subset X \setminus V_T$. Due to the definition
of $\weicnf{n}{p}{k, d}$, the random variable $N_x$ follows
the binomial distribution $Bin(\overline{p}, m)$ where
$\overline{p} = \frac{1}{2^d - 1}\frac{c\log n}{n^{d - 1}}$ and $m = \binom{n - \log n}{d - 1}$.

Write $\alpha = \frac{c}{(2^d - 1)(d - 1)!}$.
By the Chernoff bound (see Appendix B), we have
\begin{eqnarray}
\probab{N_x < k}{} &\leq& 2e^{-\frac{(\overline{p}m - k)^2}{3\overline{p}m}}  
     \leq O(k) e^{- \frac{\alpha}{3}\log n} \nonumber \\
&\in& O(n^{-\delta}) \ \ \ \ \ (\textrm{ where } 0 < \delta < \frac{\alpha}{3}).
\end{eqnarray}
Let $\mathcal{D}$ be the event that in the random formula $\mathcal{F}$, there
are two clauses  
$$
\left\{
\begin{array}{l}
\bnot{x}\bor y_{11} \bor \cdots \bor y_{1(d - 1)}, \textrm{ and} \\
\bnot{x}\bor y_{12}\bor \cdots \bor y_{2(d - 1)} 
\end{array} \right.
$$
such that
$\{y_{11}, \cdots, y_{1(d - 1)}\} \cap \{y_{12}, \cdots, y_{2(d - 1)}\} \not= \emptyset$.
The total number of such pairs of clauses is at most
$$
  (d - 1) \binom{n - \log n}{d - 1}\binom{n - \log n}{d - 2}.
$$
The probability for a specific pair to be in the random formula is
$$
   \left(\frac{1}{2^d - 1}\frac{c\log n}{n^{d - 1}} \right)^2.
$$
By Markov's inequality, we have
$$
\probab{\mathcal{D}}{}
  \in O(\frac{\log^2 n}{n}).
$$
Since the probability that the variable $x$ is not $k$-frozen is at most
$$
\probab{\{N_x < k\} \cup \mathcal{D}}{},
$$
the lemma follows. 
\end{proof}

From  Lemma \ref{lem-dsat-frozenp}, we have
\begin{lemma}
\label{lem-dsat-frozenp1}
For sufficiently large $n$,
\begin{equation}
\label{eq-conn2}
\probab{\mathcal{B}}{} < O(1) \left(n^{ - \delta} \right)^{\log n}
\end{equation}
for some $0 < \delta < \min(\frac{c}{3(2^d - 1)(d - 1)!}, 1)$.
\end{lemma}
\begin{proof}
Let $E_x$ be the event that a variable $x \in V_T$ is not $k$-frozen with respect
to $X \setminus V_T$.
Since $|V_T| = \log n$, the bound obtained in Lemma \ref{lem-dsat-frozenp}
applies to $W = X\setminus V_T$. Since for any $x \in V_T$, the event
$E_x$ only depends on the existence of clauses of the form
$$
 \bnot{x} \bor y_{i1} \bor \cdots \bor y_{i(d-1)}
$$
with $\{y_{i1}, \cdots, y_{i(d - 1)}\} \subset X\setminus V_T$, we see that
the collection of the events $\{E_x, x \in V_T\}$ are mutually independent. The Lemma follows from
Lemma \ref{lem-dsat-frozenp}.
\end{proof}
 
Next, we have the following bound on the probability $\probab{\mathcal{A}}{}$.
\begin{lemma}
\label{lem-dsat-conn}
$$
\probab{\mathcal{A}}{} \leq O(1) (\log n)^{\log n} n^{-\log n}.
$$
\end{lemma}
\begin{proof}
Recall that $\mathcal{A}$ is the event that a random instance of $\weicnf{n}{p}{k,d}$
induces all the edges of a fixed tree $T$ with vertex set $V_T$ of size $\log n$.
We follow the approach developed in 
\cite{coja-aoa-07,flaxman-soda-03,krivelevich-soda-06} and extend
the counting argument from 3-clauses to the general case of $d$-clauses with $d > 2$.

Let $F_T$ be a set of clauses such that every edge of $T$ is induced by some clause in $F_T$.
We say that $F_T$ is \textit{minimal} if deleting any clause from it leaves
at least one edge of $T$ uncovered.

Consider the different ways in which we can cover the edges of $T$ by clauses.
Treat the clauses in $F_T$ as being grouped into $d - 1$ different
groups $\{S_i, 1\leq i \leq (d - 1)\}$. A clause in the group $S_i$ is
in charge of covering exactly $i$ edges of $T$.
Note that a clause in the group $S_i$ may ``accidently" cover other edges
that are not its responsibility.  As long as each clause has its own dedicated set of
edges to cover, there won't be any risk of under-counting.

Let $s_i = |S_i|, 1\leq i\leq d - 1$. We see that $0\leq s_i \leq \log n / i$.
Since each clause in $S_i$ is dedicated to $i$ edges and there are
in total $\log n - 1$ edges, we have
\begin{equation}
\label{eq-edgecover-1}
  \sum\limits_{i = 1}^{d - 1} i s_i = \log n - 1.
\end{equation}
Counting very crudely, there are at most $\binom{\log n}{i}^{s_i}$
ways to pick the dedicated sets of  $i$ edges for the $s_i$ clauses in group
$S_i$. Since $T$ is a tree, for each set of $i$ edges
there are at most $\binom{n}{d - (i + 1)} (2^{d} - 1)$ ways to select the corresponding
clauses.  Therefore, by Markov's inequality, we have that $\probab{\mathcal{A}}{}$
can be upper bounded by
\begin{eqnarray*}
&&\sum\limits_{0\leq s_i\leq \log n} \left[
     (\log n)^{\sum\limits_{i}is_i} (2^d - 1)^{\sum\limits_{i}(d - i - 1)s_i} \right. 
        \\
&&\ \ \ \ \ \ \ \ \ \ \ \ \ \ \ \ \ \ \ \ \ \ \ \ 
   n^{\sum\limits_{i}{(d - i - 1)s_i}}
   \left.  \left(\frac{c\log n}{2^d - 1} \frac{1}{n^{d - 1}}\right)^{\sum\limits_{i}s_i}\right]
    \\
&& < O(1) \sum\limits_{0\leq s_i\leq \log n}
       (\log n)^{\log n} n^{\sum\limits_{i}(-is_i)}, \\
\end{eqnarray*}
and due to Equation (\ref{eq-edgecover-1}), we have
\begin{eqnarray*}
\probab{\mathcal{A}}{}&\leq& O(1) (\log n)^d (\log n)^{\log n} n^{-\log n + 1} \\
  &\leq& O(1) (\log n)^{2\log n} n^{-\log n}.
\end{eqnarray*}
This proves Lemma \ref{lem-dsat-conn}.
\end{proof}

Continuing the proof of Proposition \ref{prop-conn-comp}, we
combine Lemma \ref{lem-dsat-frozenp1} and Lemma \ref{lem-dsat-conn} to get
$$
 \probab{\mathcal{A} \cap \mathcal{B}}{}
  \leq O(1) (\log n)^{2\log n} n^{-\log n}
   \left(n^{ - \delta} \right)^{\log n}.
$$
Since the total number of trees of size $\log n$ is at most $n^{\log n}(\log n)^{\log n - 2}$, the probability that the residual graph of $\mathcal{F}_V$ contains a tree of size $\log n$ is
\begin{eqnarray}
\label{eq-conn-1}
&&n^{\log n}(\log n)^{\log n - 2} \probab{\mathcal{A} \cap \mathcal{B}}{} \nonumber \\
&<& O(1) (\log n)^{3 \log n}
   \left(n^{ - \delta} \right)^{\log n}
\end{eqnarray}
Proposition \ref{prop-conn-comp} follows.
\end{proof}

\ \ 

\noindent
\begin{proof}[\textbf{Proof of Theorem \ref{th-wei2cnf-sat}}]
To use Proposition \ref{prop-conn-comp} to prove that the algorithm W-SAT succeeds with high
probability, we note that the reduced formula $\mathcal{F}'$
in Line 2 of the algorithm W-SAT is sparser than the induced formula
$\mathcal{F}_V$. In fact, it is easy to see that $\mathcal{F}'$ is an induced sub-formula of
$\mathcal{F}_V$ over the set of variables that have not been assigned by the subroutine
REDUCE(). Therefore by Proposition \ref{prop-conn-comp}, with high probability $\mathcal{F}'$ decomposes into a collection of connected components, each of size at most $\log n$. It follows
that W-SAT succeeds with high probability.

Combining all the above, we conclude that the algorithm W-SAT is a fixed-paramter algorithm and
succeeds with high probability on random instances of $\weicnf{n}{p}{k, d}$.  This proves
Theorem \ref{th-wei2cnf-sat}. 
\end{proof}

\section{Discussions}
The results presented in this paper,  together with our previous
results on the threshold behavior and the resolution complexity of unsatisfiable instances
of $\weicnf{n}{p}{d, k}$ in \cite{yong08aaai},
provides a first probabilistic analysis of W[1]-complete problems.  
For WEIGHTED 2-SAT and MINI-WEIGHTED 2-SAT, the behavior 
of random instances from the studied instance distribution is fully characterized.
For WEIGHTED d-SAT with $d > 2$, the characterization is almost complete except for
a small range of the probability parameter where the parametric resolution complexity is missing. In summary, random instances of WEIGHTED d-SAT
from the random model under consideration are ``typically" fixed-parameter tractable, and
hard instances (in the sense of fixed-parameter tractability) are expected
only for MINI-WEIGHTED d-SAT.

While we believe the random model $\weicnf{n}{p}{k,d}(d')$ is very natural,
we feel that it is challenging  to come up with any alternative and natural instance distributions for weighted $d$-CNF satisfiability that are interesting and hard in terms of
the complexity of typical instances. 

On the other hand, there are still many interesting questions with the model
$\weicnf{n}{p}{k,d}$. First, the behavior of random instances of MINI-WEIGHTED d-SAT
with $d > 2$ is interesting  due to the relation between such parameterized problems
and the exponential time hypothesis of the satisfiability problem. Second,
for $p = \frac{c\log n}{n^{d - 1}}$ with $c$ small enough,
there will be sufficient number of ``isolated" variables and by simply setting $k$ of these
variables to TRUE and the rest of the variables to FALSE, we obtain a weight-$k$ satisfying
assignment. 
It is interesting to see what will happen if these isolated variables have been removed. 

\bibliographystyle{plain}
\bibliography{isaac.bib}

\begin{thebibliography}{10}

\bibitem{achlioptas01sharp}
D.~Achlioptas, P.~Beame, and M.~Molloy.
\newblock A sharp threshold in proof complexity.
\newblock In {\em Proceedings of STOC'01}, pages 337--346, 2001.

\bibitem{alon97siam}
N.~Alon and N.~Kahale.
\newblock A spectral technique for coloring random 3-colorable graphs.
\newblock {\em SIAM J. Computing}, 26:1733--1748, 1997.

\bibitem{beame02efficiency}
P.~Beame, R.~Karp, T.~Pitassi, and M.~Saks.
\newblock The efficiency of resolution and {D}avis-{P}utnam procedures.
\newblock {\em SIAM J. on Computing}, 31(4):1048--1075, 2002.

\bibitem{bollobas01}
B.~Bollobas.
\newblock {\em Random Graphs}.
\newblock Cambridge University Press, 2001.

\bibitem{cheeseman91}
P.~Cheeseman, B.~Kanefsky, and W.~Taylor.
\newblock Where the {\em really} hard problems are.
\newblock In {\em Proceedings of the 12th International Joint Conference on
  Artificial Intelligence}, pages 331--337. Morgan Kaufmann, 1991.

\bibitem{coja-aoa-07}
A.~Coja{-}Oghlan, M.~Krivelevich, and D.~Vilenchik.
\newblock Why almost all satisfiable k-{CNF} formulas are easy.
\newblock In {\em Proc. of the 13th International Conference on Analysis of
  Algorithms}, pages 89--102, 2007.

\bibitem{cook97}
S.~Cook and D.~Mitchell.
\newblock Finding hard instances of the satisfiability problem: A survey.
\newblock In Du, Gu, and Pardalos, editors, {\em Satisfiability Problem: Theory
  and Applications}, volume~35 of {\em DIMACS Series in Discrete Mathematics
  and Theoretical Computer Science}. American Mathematical Society, 1997.

\bibitem{downey99book}
R.~Downey and M.~Fellows.
\newblock {\em Parameterized Complexity}.
\newblock Springer, 1999.

\bibitem{feige01semi}
U.~Feige and J.~Kilian.
\newblock Heuristics for semirandom graph problems.
\newblock {\em J. of Computer Science and Systems}, 63:639--671, 2001.

\bibitem{flaxman-soda-03}
A.~Flaxman.
\newblock A spectral technique for random satisfiable 3{CNF} formulas.
\newblock In {\em Proc. of 14th ACM-SIAM Symposium on Discrete Algorithms},
  pages 357--363, 2003.

\bibitem{yong08aaai}
Y.~Gao.
\newblock Phase transitions and complexity of weighted satisfiability and other
  intractable parameterized problems.
\newblock In {\em Proceedings of the 23rd AAAI Conference on Artificial
  Intelligence (AAAI'08), to appear}, 2008.

\bibitem{gao07jair}
Y.~Gao and J.~Culberson.
\newblock Consistency and random constraint satisfaction models.
\newblock {\em Journal of Artificial Intelligence Research}, 28:517--557, 2007.

\bibitem{krivelevich-soda-06}
M.~Krivelevich and D.~Vilenchik.
\newblock Solving random satisfiable 3{CNF} formulas in expected polynomial
  time.
\newblock In {\em Proc. of 17th ACM-SIAM Symposium on Discrete Algorithms},
  pages 454--463, 2006.

\bibitem{marx05csp}
D.~Marx.
\newblock Parameterized complexity of constraint satisfaction problems.
\newblock {\em Computational Complexity}, (2):153--183, 2005.

\bibitem{molloy02}
M.~Molloy.
\newblock Models and thresholds for random constraint satisfaction problems.
\newblock In {\em Proceedings of STOC'02}, pages 209 -- 217. ACM Press, 2002.

\bibitem{rolf06book}
R.~Neidermeier.
\newblock {\em Invitation to Fixed-Parameter Algorithms}.
\newblock Oxford University Press, 2006.

\end{thebibliography}

\newpage

\section{Appendix A - Generalization to the Model $\weicnf{n}{p}{k, d}(d')$}
\label{sec-extension}
Consider the model $\weicnf{n}{p}{k, d}(d'), d' < d,$ that generalizes
the model $\weicnf{n}{p}{k, d}$.  To generate a random instance $\mathcal{F}$
of $\weicnf{n}{p}{k, d}(d')$,  we first construct a random hypergraph
$\mathcal{G}(n, p, d)$ in the same way as with the random model $\weicnf{n}{p}{k, d}$.
For each hyperedge $\{x_{i_1}, \cdots, x_{i_d}\}$, we include in $\mathcal{F}$
a $d$-clause selected uniformly at random from the set of the $d$-clauses
over  $\{x_{i_1}, \cdots, x_{i_d}\}$ that contain at least $d'$ negated literals.

Note that with the above definition, the original model $\weicnf{n}{p}{k, d}$
is just $\weicnf{n}{p}{k, d}(1)$.
Similar to the analysis for $\weicnf{n}{p}{k, d}$ presented in \cite{yong08aaai}, 
the following threshold behavior of the solution probability can be established  
\begin{lemma} 
Consider a random instance $(\weicnf{n}{p}{k,d}(d'), k)$ of WEIGHTED d-SAT.
Let  $p = \frac{c\log n}{n^{d - d'}}$ with $c > 0$ being a constant and
let $c^{*} = a_d(d - d')!$ with $a_d$ being the number of $d$-clauses over a fixed
set of $d$ variables that contain at least $d'$ negated literals. We have
\begin{eqnarray}
&&\lim\limits_{n}\probab{ \weicnf{n}{p}{k, d}(d') \textrm{ is satisfiable } }{}
  = \left\{\begin{array}{ll}
             1, & \mbox{if }  c < c^{*}, \\
             0, & \mbox{if }  c > c^{*}
            \end{array}
    \right. \nonumber
\end{eqnarray}
\end{lemma}

For  $p = \frac{c\log n}{n^{d - d'}}$,
the algorithm W-SAT can be adapted to solve  a random instance of 
$\weicnf{n}{p}{k, d}(d')$ in $O(k^2n + n^{O(1)})n^{(d' - 1)}$ time by using
the following generalization of a $k$-frozen variable:

\begin{definition}
Let $(\mathcal{F}, k)$ be an instance of WEIGHTED d-SAT where $\mathcal{F}$ is a d-CNF formula
and $k$ is the parameter. Let $2 \leq d' \leq d $ be a fixed integer. 

Consider a variable $x$, a set of $(d' - 1)$ variable $S = \{x_1, \cdots, x_{d' - 1}\}$, and
a collection of subsets of variables $\mathcal{Y} = \{Y_i, 1\leq i\leq k\}$ where
$$
  Y_i = \{y_{ij}, 1\leq j \leq (d - d') \}
$$
is a subset of $X \setminus (\{x\} \cup S)$.
We say that the collection $\mathcal{Y}$  of subsets of variables
\textbf{freeze} $x$ \textbf{on $S$} if
\begin{enumerate}
\item  $Y_i \cap Y_j = \phi, \forall i, j$.
\item for each $1\leq i\leq k$, the clause
\begin{eqnarray*}
 \bnot{x}_1 \bor \cdots \bor \bnot{x}_{d' - 1} \bor 
     \bnot{x} \bor y_{i1} \bor \cdots \bor y_{i(d - d')}
\end{eqnarray*}
is in the formula $\mathcal{F}$.
\end{enumerate}
\end{definition}

\begin{lemma}
If $x$ is $k$-frozen on $S = \{x_1, \cdots, x_{d' - 1 }\}$, then
assigning all the variables in $S$ to TRUE forces $x$ to be FALSE.
\end{lemma}

The modification of W-SAT  to solve random instances of 
$\weicnf{n}{p}{k, d}(d')$ is as follows: For each of 
the $\binom{n}{d' - 1}$ possible sets of $(d' - 1)$ 
variables $S = (x_1, \cdots, x_{d' - 1})$, set  them to TRUE and all the
variables that are $k$-frozen on $S$ to FALSE; Apply the subroutine $REDUCE()$ to
obtain a reduced formula $\mathcal{F}'$; Use the same technique in W-SAT
to check to see if $\mathcal{F}'$ has a satisfying assignment of weight
$k - (d' - 1)$. The overall running time is $O(k^2n + n^{O(1)})n^{(d' - 1)}$.     

\section{Appendix B - Random Instances of MINI-WEIGHTED $d$-SAT}
In the proof in Section 4 and in this section, we use the following Chernoff bound
\begin{lemma}
\label{lem-chernoff}
Let $I$ be a binomial random variable with expectation $\mu$. We have
$$
 \probab{|I - \mu| > t}{} \leq 2e^{-\frac{t^2}{3\mu}}.
$$
\end{lemma}

As a variant of WEIGHTED $d$-SAT, the problem MINI-WEIGHTED $d$-SAT with parameter $k$ asks  
if for a given $d$-CNF formula, there 
is a satisfying assignment of weight $k\log n$. For random $d$-CNF formula 
$\weicnf{n}{p}{k,d}$, the algorithm W-SAT for MINI-WEIGHTED $d$-SAT needs to be adapted to  
make use of
the existence of $k\log n$-frozen variables. To guarantee that W-SAT  still succeeds with
high probability, a result similar to Proposition \ref{prop-conn-comp} is needed.
This amounts to showing that the probability for a variable $x$ to be $k\log n$-frozen is small enough. For
$p = \frac{c\log n}{n^{d - 1}}$ with $c > k2^{d - 1}(d - 1)!$, this is the case.
\begin{theorem}
\label{col-mini-sat}
There is an $O(k^2n + n^{O(1)})$-time algorithm that solves with high probability
a random instance $(\weicnf{n}{p}{k,d}, k)$ of
MINI-WEIGHTED $d$-SAT for any $p = \frac{c\log n}{n^{d - 1}}$ with $c > k(2^d - 1)(d - 1)!$.
\end{theorem}
\begin{proof}
The proof is almost the same as the proof of Proposition \ref{prop-conn-comp} except that
we need to establish  an upper bound on the probability that a variable is not
$k\log n$-frozen. For  $c > k(2^d - 1)(d - 1)!$,
Lemma \ref{lem-chernoff} on the tail probability of a binomial random variable
is still effective 
and the arguments made in the second half of the proof of Lemma
\ref{lem-dsat-frozenp} and in the proof of Lemma \ref{lem-dsat-frozenp1} are still valid.
The only difference is the accuracy of the upper bound.
In this case, we have $\probab{\mathcal{B}}{} \leq O(1) \frac{1}{n^{\delta\log n}}$
where $0 < \delta < \min(\frac{(k - c)^2c}{3(2^d - 1)(d - 1)!}, 1)$, and
this is sufficient for the result to hold.
\end{proof}
 
\end{document}